\documentclass[letterpaper]{article} 
\usepackage{aaai23}  
\usepackage{times}  
\usepackage{helvet}  
\usepackage{courier}  
\usepackage[hyphens]{url}  
\usepackage{graphicx} 
\usepackage{natbib}  
\usepackage{caption} 
\usepackage{algorithm}
\usepackage{algorithmic}

\usepackage{newfloat}
\usepackage{listings}
\DeclareCaptionStyle{ruled}{labelfont=normalfont,labelsep=colon,strut=off} 
\floatstyle{ruled}
\newfloat{listing}{tb}{lst}{}
\floatname{listing}{Listing}
\usepackage{algorithm}
\usepackage{algorithmic}
\usepackage{booktabs}
\usepackage{amsthm}
\usepackage{amssymb}
\usepackage{amsmath}
\usepackage{bm}
\newtheorem{definition}{Definition}
\usepackage{multirow}
\usepackage{graphicx}

\title{Improving Zero-Shot Coordination Performance Based on Policy Similarity}
\author{
    Lebin Yu, Yunbo Qiu, Quanming Yao, Xudong Zhang, Jian Wang
}
\affiliations{
    Department of Electronic Engineering, Tsinghua University\\
    Beijing, China\\
    \{yulb19,qyb18\}@mails.tsinghua.edu.cn, quanmingyao@gmail.com, \{zhangxd, jianwang\}tsinghua.edu.cn
}

\begin{document}

\maketitle

\begin{abstract}
Over these years, multi-agent reinforcement learning has achieved remarkable performance in multi-agent planning and scheduling tasks. It typically follows the self-play setting, where agents are trained by playing with a fixed group of agents. However, in the face of zero-shot coordination, where an agent must coordinate with unseen partners, self-play agents may fail. Several methods have been proposed to handle this problem, but they either take a lot of time or lack generalizability. In this paper, we firstly reveal an important phenomenon: the zero-shot coordination performance is strongly linearly correlated with the similarity between an agent's training partner and testing partner. Inspired by it, we put forward a \underline{S}imilarity-\underline{B}ased \underline{R}obust \underline{T}raining (SBRT) scheme that improves agents' zero-shot coordination performance by disturbing their partners' actions during training according to a pre-defined policy similarity value. To validate its effectiveness, we apply our scheme to three multi-agent reinforcement learning frameworks and achieve better performance compared with previous methods. 
\end{abstract}

\section{Introduction}
In recent years, multi-agent reinforcement learning has been attracting increasing attention for its broad applications in automated planning and scheduling tasks such as robot navigation~\cite{li2019disjoint}, traffic control~\cite{calvo2018heterogeneous, ghosh2021deep} and fleet management~\cite{lin2018efficient, bogh2022distributed}. Unlike single-agent tasks, where the agent only needs to interact with a stationary environment, multi-agent setting brings instability caused by agents' continuous updating of their policies~\cite{lowe2017multi, mahajan2019maven}. To solve this problem, researchers let a fixed set of agents train and test together, which is called self-play~\cite{tesauro1994td}. Self-play makes it easy for agents to 
quickly understand partners' policies and adjust their own plans. However, self-play-trained agents usually cannot cooperate with unseen agents well during testing \cite{hu2020other}. 

In order to deeply study the generalization performance of cooperative agents, researchers put forward the concept of zero-shot coordination \cite{hu2020other}, where agents must coordinate with unfamiliar partners that they have not seen before. Since then, several methods have been proposed to solve the problem, which can be divided into two categories. The first is population-based training \cite{shih2021critical,lucas2022any}, where agents are trained with multiple and diverse partners. This kind of method allows agents to learn more generalized policies, but is time-consuming. The second is avoiding specific conventions among self-play agents by making agents reason about the task \cite{hu2021off,cui2021k} or breaking symmetries \cite{hu2020other,treutlein2021new}. Agents trained by this kind of framework can cooperate well with unseen partners trained by the same framework, but perform poorly when cooperating with partners trained by other frameworks. 

In these papers, cross-play is adopted to evaluate zero-shot coordination performance, where individually trained agents are required to play with each other. However, experiments show that the cooperative performance of different combinations of agents varies widely, even if they are trained by the same framework with different random number seeds. We give an intuitive explanation for this phenomenon: If the testing partner is similar to an agent's training partner, the agent might cooperate with this unseen partner well. Since the similarity between agents varies widely, so does the cross-play score. To measure the similarity, we define \underline{C}onditional \underline{P}olicy \underline{S}imilarity between an agent's \underline{T}raining partner and \underline{T}esting partner (CPSTT), which is the expected probability of an agent's training partner and testing partner taking the same action when coordinating with the agent. In addition, we conduct numerous experiments in a cooperative card game Hanabi~\cite{bard2020hanabi} and reveal a strong linear correlation between CPSTT and cross-play scores.   

Inspired by this linear correlation, we propose a light-weighted scheme, \underline{S}imilarity-\underline{B}ased \underline{R}obust \underline{T}raining (SBRT), to improve the zero-shot coordination performance of agents. It randomly disturbs the training partners' policies during training according to a pre-defined CPSTT value, in which way the overfitting of the agent to specific training partners is alleviated. We apply SBRT to three multi-agent reinforcement learning frameworks,  IQL~\cite{tan1993multi}, VDN~\cite{sunehag2018value} and SAD~\cite{hu2019simplified}, and experiments confirm that their zero-shot coordination performance is improved.

Our contributions are summarized below:
\begin{enumerate}
    \item We define CPSTT and conduct extensive experiments to reveal its strong linear correlation with cross-play scores: cross-play scores increase almost linearly with it. This finding can provide a new perspective on zero-shot coordination.
    \item Inspired by the linear correlation, we propose a light-weighted scheme SBRT that randomly perturbs agents' partners' policies during training according to a pre-defined CPSTT value. 
    \item We apply SBRT to three multi-agent reinforcement learning frameworks and experimentally confirm the effectiveness of our methods.
\end{enumerate}

\section{Background}
\subsection{Decentralized Partially Observable Markov Decision Process}
We use a decentralized partially observable Markov decision process  (Dec-POMDP) ~\cite{nair2003taming} to model a multi-agent cooperation task. The environment is partially observable with $N$ agents in it. At timestep $t$, with the global environment state being $s_t \in \mathbf{S}$, agent $i$ gets its observation $o_t^i \sim O(o|i,s_t)$ and chooses an action $a_t^i$ according to its action-observation trajectory $\tau_t^i=\{o_0^i,a_0^i,...,o_t^i \}$ as well as its policy $\pi^i$: $a_t^i \sim \pi^i(a^i|\tau_t^i)$. The environment then feeds back a shared reward $r_t = R(s_t, \vec{a}_t)$ and updates the global state $s_{t+1} \sim T(s'|s_t, \vec{a}_t)$, where $\vec{a}_t = [a_t^1,...,a_t^N]$ is the joint action. The training goal is to maximize the expected return $J(\pi^1,...,\pi^N)=\mathbb{E}_{\pi^1,...,\pi^N}[\sum_t \gamma^t r_t]$.

To accomplish the training goal, researchers have proposed many multi-agent reinforcement learning frameworks, including policy-gradient-based ones~\cite{foerster2018counterfactual, kuba2021trust}  and value-based ones~\cite{rashid2018qmix, son2019qtran}. They typically make a fix group of agents play and learn together so that agents can quickly learn cooperative behaviors. 

\subsection{Zero-shot Coordination and Cross-play}
Following the common setting in this area~\cite{hu2020other}, we formulate zero-shot coordination in two-agent scenarios. Suppose an agent with policy $\pi$ is trained along with a partner agent with policy $\pi_p$, the training goal is described as follows:
\begin{equation}
\label{eqa:basic_goal}
   \pi^*, \pi_p^* = \arg\max_{\pi,\pi_p} J(\pi,\pi_p)
\end{equation}

If $\pi$ and $\pi_p$ continue to cooperate during testing, they are likely to get high scores. However, zero-shot coordination requires agents to coordinate with unseen partners. A similar setting is ad-hoc teamwork~\cite{rahman2021towards, zand2022fly}, where agents must model unfamiliar partners' policies and adjust their own strategies during interaction. In comparison, zero-shot coordination does not allow agents to change their policies when facing unseen partners.

To evaluate agents' performance in this setting, researchers commonly conduct cross-play, where individually trained agents are put together to cooperate during testing. Besides, the zero-shot coordination performance of a training framework is measured by the average cross-play score of multiple agents trained with this framework. Below we give a formulaic representation of this process. Suppose the policy trained by a training framework $M$ has a distribution $P_M$, then in the case that the testing partner is trained by a training framework $M_o$, the zero-shot coordination performance of $M$ can be expressed as:
\begin{equation}
   Z(M) = \mathbb{E}_{\pi \sim P_M, \pi_o \sim P_{M_o}}[J(\pi,\pi_o)]
\end{equation}
It is evident that $Z(M)$ heavily relies on $M_o$. Based on the different choices of $M_o$, cross-play can be divided into two categories:

\textbf{(1)Intra-algorithm cross-play:} $M_o = M$, which means the unseen partners are trained with the same training framework but different random number seeds~\cite{cui2021k}. This kind of test is easily accessible and objective, but its disadvantage is also obvious: it does not indicate whether the agents can cooperate well with agents obtained by other frameworks. 

\textbf{(2)Inter-algorithm cross-play:} $M_o = (M_1+M_2+...+M_K)$, where $M_1,M_2,...,M_K$ are other multi-agent reinforcement learning frameworks~\cite{lucas2022any}. This seems to be more comprehensive, however, the K tested frameworks may not represent all feasible frameworks, so the results may be biased.

To evaluate the zero-shot coordination performance of agents more comprehensively, we conduct both intra-algorithm and inter-algorithm cross-play in the experiments.

\begin{figure*}[htbp]
   \begin{center}
   \begin{minipage}[t]{0.31\textwidth}
   \centering
   \includegraphics[width=1\textwidth]{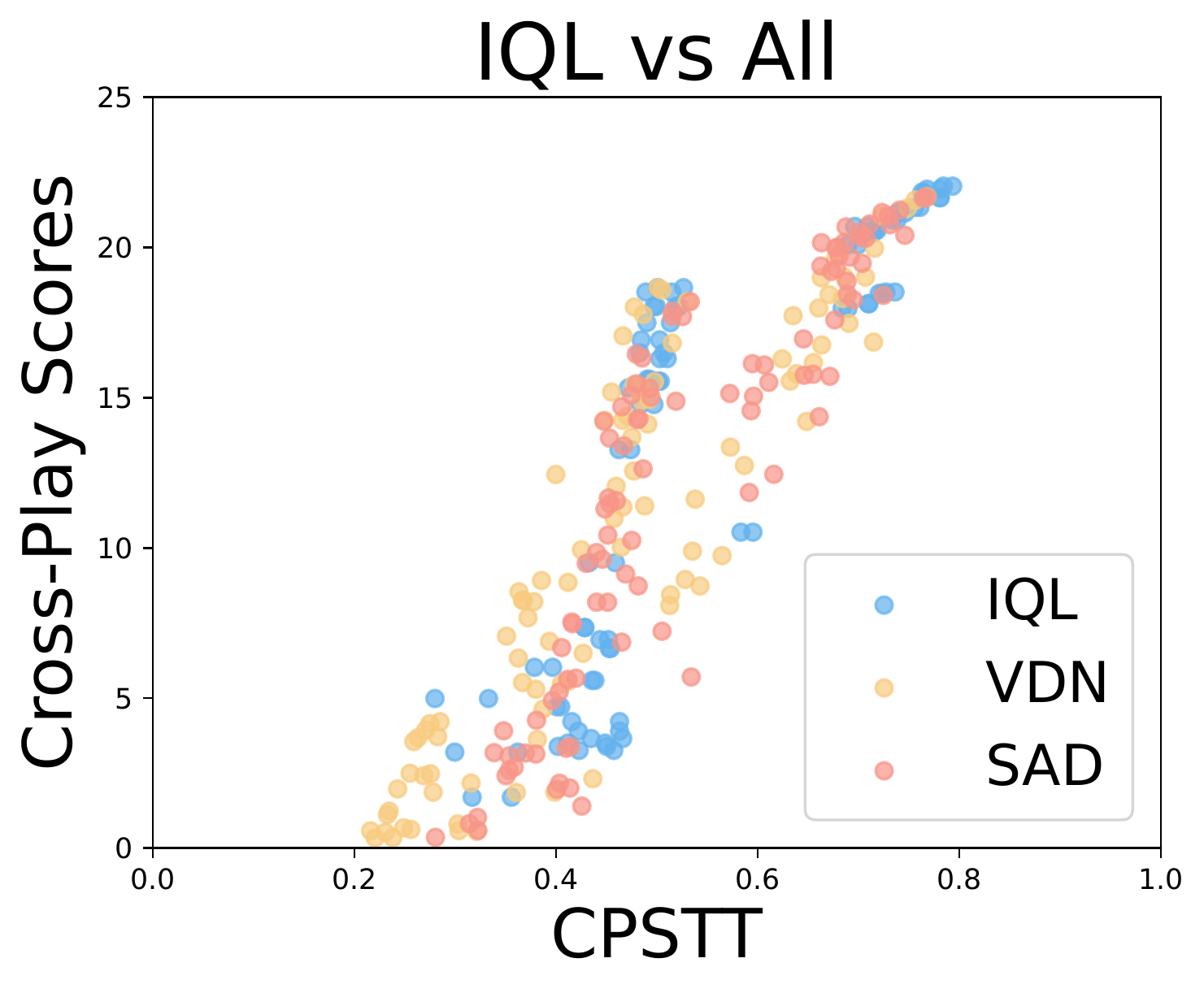}
   \centerline{$r_p$=0.874}
   \end{minipage}
   \begin{minipage}[t]{0.31\textwidth}
   \centering
   \includegraphics[width=1\textwidth]{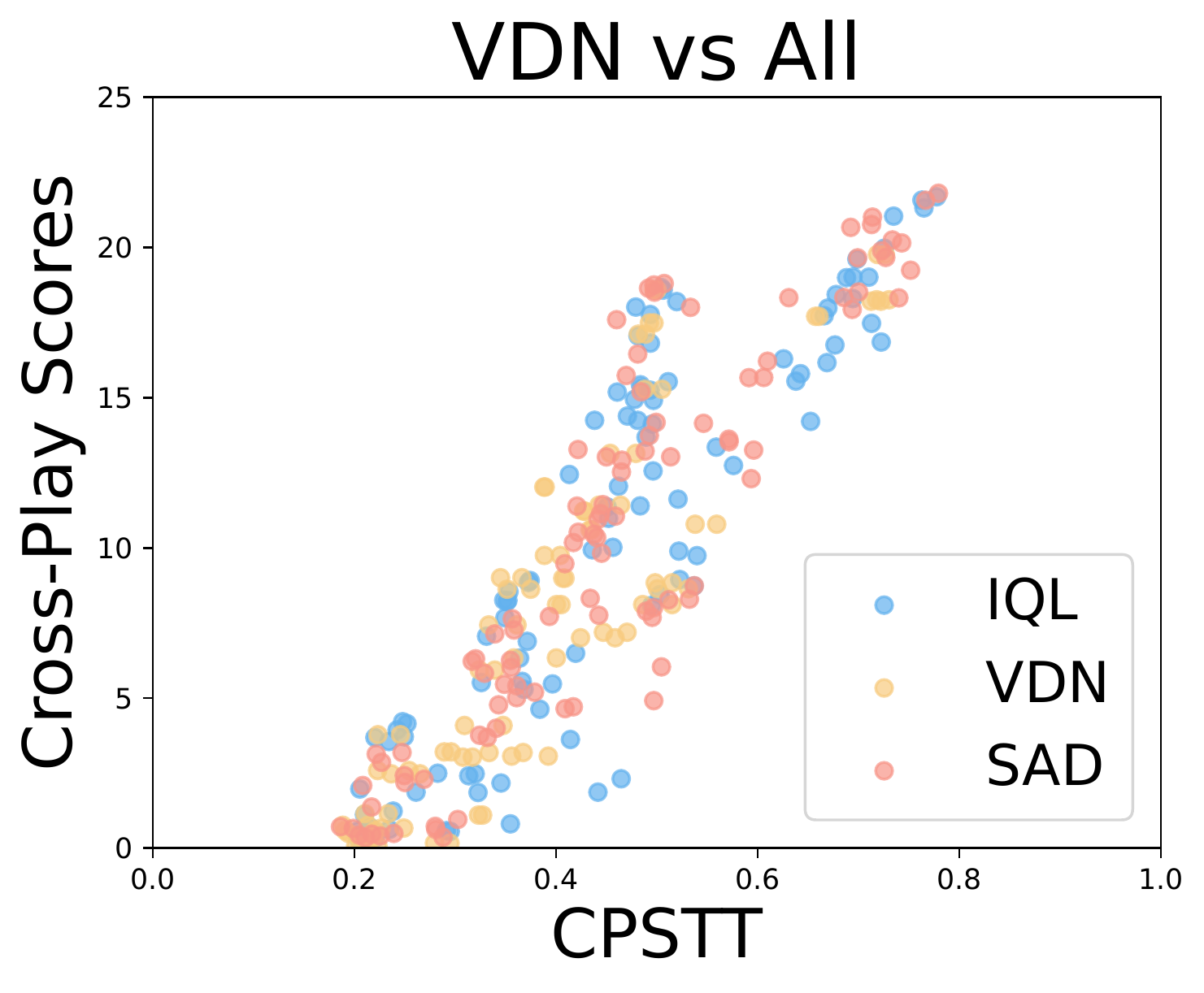}
   \centerline{$r_p$=0.905}
   \end{minipage}
   \begin{minipage}[t]{0.31\textwidth}
   \centering
   \includegraphics[width=1\textwidth]{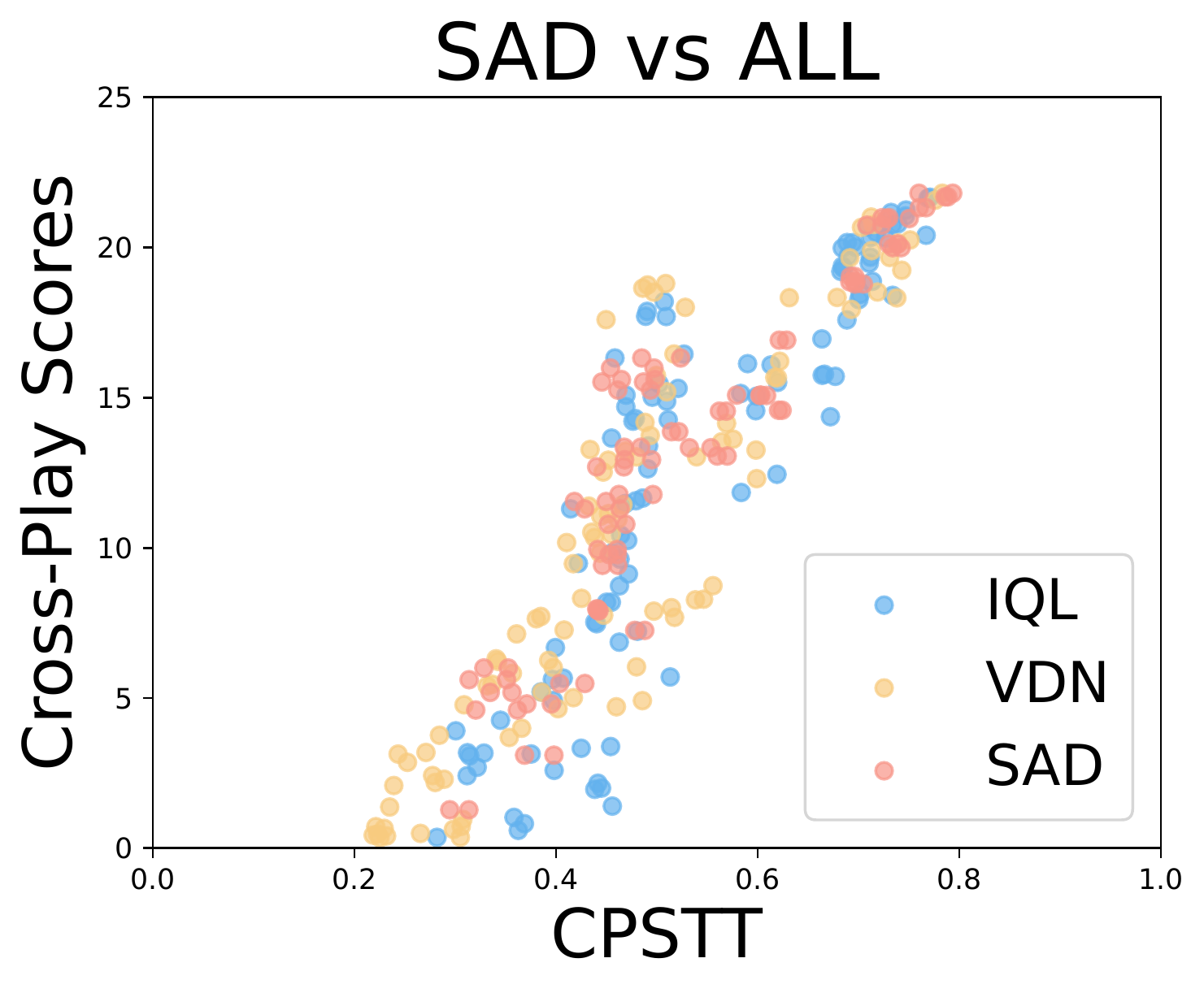}
   \centerline{$r_p$=0.904}
   \end{minipage}
   \end{center}
   \caption{Each figure shows the detailed cross-play results of one kind of model with all kinds of models. $r_p$ is Pearson correlation coefficient, a statistic describing the linear correlation between two factors ($r_p=1$ represents a completely linear relationship). Generally speaking, the linear correlation between CPSTT and cross-play scores is strong.}
   \label{fig:base_test}
   \end{figure*}
\section{Conditional Policy Similarity}
Cross-play scores of different combinations of agents vary a lot, even for the combinations of agents trained by the same framework. To account for this phenomenon, we make an intuitive guess that an agent should coordinate with an unseen partner well if this partner is similar to the agent's training partner. In this section, we define CPSTT to measure this similarity, propose a way to estimate it, and present experimental results that indicate the strong linear correlation between it and cross-play scores. 

\subsection{Definition and Estimation of Conditional Policy Similarity}
Below we give our definition of conditional policy similarity:
\begin{definition}
   In a two-agent game, the conditional policy similarity between $\pi_1$ and $\pi_2$ conditioned on $\pi$ is:
   \begin{equation}
      S_{\pi}(\pi_1,\pi_2) = \mathbb{E}_{\tau \sim P_{\tau(\pi,\pi_1)}}[\pi_1(\tau)=\pi_2(\tau)]
   \end{equation}
   where $P_{\tau(\pi,\pi_1)}$ denotes the distribution of action-observation trajectory generated by $\pi$ and $\pi_1$ playing with each other.
\end{definition}
Conditional policy similarity measures how similar $\pi_2$ is to $\pi_1$ from $\pi$'s perspective, and can be estimated in a Monte Carlo approach: Let $\pi$ and $\pi_1$ play the game several times, and assume there are total $n$ timesteps. Then, $\pi_1$ makes $n$ decisions $\{\pi_1(\tau_1),\pi_1(\tau_2),...,\pi_1(\tau_n)\}$ based on $n$ action-observation trajectories: $\{\tau_1,\tau_2,...,\tau_n\}$. Let $\pi_2$ acts based on these $n$ trajectories, and then the estimate for $S_{\pi}(\pi_1,\pi_2)$ becomes:
\begin{equation}
   \label{eqa:estimate_cps}
   \bar{S_{\pi}}(\pi_1,\pi_2) = \frac{1}{n}\sum_{i=1}^n \bm{1}_\mathrm{\pi_1(\tau_i)=\pi_2(\tau_i)}
\end{equation}
In this paper, we focus on the conditional policy similarity between an agent's training partner and testing partner (i.e. CPSTT) in zero-shot coordination. Given an agent with policy $\pi$, its training partner policy $\pi_p$ and testing partner policy $\pi_o$, CPSTT$=S_{\pi}(\pi_o,\pi_p)$. 

\subsection{Testbed: Hanabi}
We experiment on Hanabi \cite{bard2020hanabi}, a cooperative card game often used to study zero-shot coordination. Each card has a color and a rank, and players are required to play cards in a legal order to complete five stacks of cards, one for each color. There are 5 colors and 5 ranks and the maximum score is 25. Hanabi has a typical cooperative reward scheme, where agents get rewarded for increasing team scores and punished for losing team life tokens. Note that each player can see the cards in everyone's hand except its own, hence it must guess what their cards are based on others' cues as well as provide valuable information for others. 

Since this task requires agents to reason about the beliefs and intentions of partners and self-play agents are quite familiar with their training partners, they can achieve super-human performance \cite{hu2019simplified}. However, they easily learn special conventions to get high self-play scores. For example, one can hint `blue' to let a partner play the first card from the left. This kind of trick helps quickly achieve self-play goals in (\ref{eqa:basic_goal}) but does not help agents cooperate well with unfamiliar partners. As a result, zero-shot coordination is difficult in Hanabi, and a large part of the work in this field experiments on it~\cite{hu2020other, treutlein2021new, lucas2022any, hu2021off}.

\subsection{Explore the Relationship Between CPSTT and Cross-play Scores}
\label{sec:maintest}
We conduct abundant experiments to see whether CPSTT and cross-play scores are correlated. In specific, we experiment on the three types of models: IQL, VDN and SAD. They are typical self-play algorithms, and we apply parameter sharing to them to accelerate training, which is a widely-used technique~\cite{christianos2020shared} for self-play in homogeneous multi-agent games. Therefore, the agent's training partner is itself and CPSTT for these four kinds of models is $S_{\pi}(\pi_o,\pi)$. We train 10 models of each framework with different seeds and pair these 30 models to get 900 combinations. For each combination, we record the average scores of 10000 games and estimate $S_{\pi}(\pi_o,\pi)$ according to (\ref{eqa:estimate_cps}). 

To visualize how conditional policy similarity affects cross-play scores, we exhibit the detailed cross-play results in Fig.~\ref{fig:base_test}. These figures show the same pattern: scores increase with similarities. This confirms our guess that if the testing partner is similar to an agent's training partner, the cooperation tends to be good. Then how strong their linear correlation is exactly? We present the Pearson correlation coefficient ($r_p$) beneath each figure. The lowest is 0.874 (IQL), and the highest is 0.905 (SAD). For reference, $r_p=1$ represents an unrealistically perfect correlation. Therefore, the linear correlation between CPSTT and cross-play scores is strong. Based on this fact, we can make a more comprehensive analysis of zero-shot coordination as well as propose new ways to increase cross-play scores.

\section{Improving Cross-play Scores Based on Conditional Policy Similarity}

Our experiments in the above section show that cross-play scores increase almost linearly with CPSTT. Therefore, cross-play scores can be increased in two ways based on this correlation. The first is to increase CPSTT. However, zero-shot coordination means cooperating with unseen partners, and it is almost impossible to increase the policy similarity between training partners and unknown agents. The second is more feasible, that is to improve cross-play scores while CPSTT is fixed, based on which we propose our scheme. 

\subsection{Similarity-Based Robust Training}

\label{sec:sbrt}
Inspired by robust reinforcement learning \cite{mankowitz2018learning}, we put forward a robust training scheme SBRT that can be applied to common self-play multi-agent reinforcement learning frameworks. The training objective of it is 
\begin{equation}
   \label{eqa:sbrt}
      \pi^*, \pi_p^* = \arg\max_{\pi,\pi_p} J(\pi,\pi_a) \ \       s.t. CPSTT_{\pi}(\pi_p,\pi_a) = \alpha
\end{equation}
In the first phase of training, we set $\alpha=1$, which means the training is the same as self-play, allowing the agents to quickly optimize their policies. After training for $N_{st}$ epochs, when the training is close to convergence, we set $\alpha=\alpha_r$ and perform robust training for $N_{rt}$ epochs. $\alpha_r, N_{st}$ and $N_{rt}$ are training hyperparameters. We implement $\pi_a$ by disturbing the partner's chosen action when it interacts with the environment. To be specific, given an action-observation trajectory $\tau$ and the action generated by $\pi_p$: $a_p=\pi_p(\tau)$, 
\begin{eqnarray}
   \pi_a(\tau)=\left\{
   \begin{aligned}
      a_p, \quad \quad \quad \quad &with \ probability \ \alpha \\
      a_{alt}\neq a_p, \quad &with \ probability \ 1-\alpha
   \end{aligned}
   \right.
\end{eqnarray}
We test three ways of choosing $a_{alt}$:

\textbf{Worst alternative action:} This choice requires $\pi_a$ to satisfy 
\begin{equation}
   \pi_a = \arg\min_{\pi_a} J(\pi,\pi_a), \quad s.t. CPSTT_{\pi}(\pi_p,\pi_a) = \alpha
\end{equation}
This method follows the idea of adversarial training~\citep{pan2019risk}, where an adversary tries to minimize the main agent's performance. Under a DQN-based framework with a Q-function $Q(\tau,a)$, $a_{alt}=\arg\min_{a}Q(\tau,a)$.

\textbf{Best alternative action:} This choice requires $\pi_a$ to satisfy 
\begin{equation}
   \pi_a = \arg\max_{\pi_a} J(\pi,\pi_a), \quad s.t. CPSTT_{\pi}(\pi_p,\pi_a) = \alpha
\end{equation}
The reason for this choice is that the policies of zero-shot coordination partners are usually not bad, hence this restriction might make $\pi_a$ more like them. Under a DQN-based framework with a q function $Q(\tau,a)$, $a_p = \arg\max_{a}Q(\tau,a)$ and $a_{alt}=\arg\max_{a \neq a_p}Q(\tau,a)$.

\textbf{Random alternative action:} This solution simply picks a random feasible action as $a_{alt}$. It maximizes the exploration of possible unseen partner policies instead. We find in experiments that this is the best solution and choose it to implement SBRT in the main experiments.

\subsection{Evaluation of SBRT}
To see whether SBRT improves the zero-shot coordination performance of common self-play agents, we combine our scheme with three baseline multi-agent reinforcement learning frameworks: IQL, VDN and SAD. We also include OP~\cite{hu2020other} in the comparison, which breaks task-specific symmetries to increase cross-play scores. In this task, OP shuffles the color of the cards to prevent agents from forming special consensus (e.g. hinting ‘blue’ to let a partner play the first card from the left).

As for the training hyperparameters, $\alpha_r=0.8$, $N_{st}=400$ and $N_{rt}=100$. Besides, we set the total training epochs of baseline and OP models to $500$ to be consistent with SBRT. To increase the reliability of the results, we train 10 models with different random number seeds for each method.  

We conduct inter-algorithm and intra-algorithm cross-play to measure the zero-shot coordination performance of the testing models. Intra-algorithm cross-play requires agents to cooperate with partners trained by the same framework. For example, IQL with IQL models, VDN+OP with VDN+OP models. Inter-algorithm cross-play requires agents to cooperate with partners trained by other frameworks. For example, for IQL, IQL+OP and IQL+SBRT models, their inter-algorithm cross-play partners are raw VDN and SAD models.

Results shown in Table~\ref{tab:main_test} confirm that that SBRT effectively increases the inter-algorithm and intra-algorithm cross-play scores of IQL, VDN and SAD. An intuitive explanation for the validity of SBRT is that it diversifies the policies of agents training partners, thereby avoiding the formation of special consensus and allowing agents to perform better when cooperating with unseen testing partners.

\begin{table}[t]
   \caption{Mean and standard error of cross-play scores of baseline, OP and SBRT models}
   \label{tab:main_test}
   \begin{center}
 \begin{tabular}{lll}
 \toprule
 \multicolumn{1}{c}{\multirow{2}[4]{*}{Model Type}} & \multicolumn{2}{c}{Cross-Play Scores} \\
\cmidrule{2-3}   & Intra-Algorithm & Inter-Algorithm \\
 \midrule
 IQL & 13.55$\pm$0.99 & 12.39$\pm$0.21 \\
 IQL+OP & 11.93$\pm$1.10 & 11.35$\pm$0.22 \\
 IQL+SBRT & \textbf{15.72$\pm$0.70} & \textbf{13.25$\pm$0.20} \\
 \midrule
 VDN & 7.70$\pm$0.86 & 9.97$\pm$0.20 \\
 VDN+OP & 6.11$\pm$0.81 & 8.65$\pm$0.22 \\
 VDN+SBRT & \textbf{11.38$\pm$0.92} & \textbf{12.14$\pm$0.20} \\
 \midrule
 SAD & 14.12$\pm$0.86 & 11.32$\pm$0.20 \\
 SAD+OP & 11.35$\pm$0.93 & 9.25$\pm$0.19 \\
 SAD+SBRT & \textbf{14.70$\pm$0.77} & \textbf{13.19$\pm$0.19} \\
 \bottomrule
 \end{tabular}%

   \end{center}
   \end{table}

\section{Conclusion and Future Work}
We focus on zero-shot coordination, where agents are required to cooperate with unseen partners. Several methods have been proposed to solve this problem, however, they either lack generalizability or take a lot of time. In this paper, we define conditional policy similarity and find it important in zero-shot coordination: cross-play scores are strongly correlated with CPSTT, and the Person correlation coefficient between them can be as high as 0.905. Based on this correlation, we propose a light-weighted scheme SBRT, that improves the zero-shot coordination performance of agents by perturbing the training partners' policies according to a predefined CPSTT value. To confirm its effectiveness empirically, we apply it to three multi-agent reinforcement learning frameworks (IQL, VDN and SAD). Results show that SBRT successfully improves the inter-algorithm and intra-algorithm cross-play scores of these models. We hope the discovery of the strong correlation between cross-play scores and CPSTT along with our SBRT scheme provide a new perspective for research in this area.

\bibliography{aaai23}

\end{document}